\documentclass[doublecol]{epl2}
\def\be{\begin{equation}}
\def\ee{\end{equation}}

\usepackage{graphicx}
\usepackage{bm}
\usepackage{amssymb}

\title{Liquid Hertz Contact: softness of weakly deformed drops on non-wetting substrates}
\author{F. Chevy\inst{1} \and A. Chepelianskii\inst{2} \and D. Qu\'er\'e\inst{3} \and E. Rapha\"el\inst{4}}
\shortauthor{F. Chevy \etal}

\institute{
  \inst{1} Laboratoire Kastler Brossel, CNRS, UPMC, \'Ecole Normale Sup\'erieure, 24 rue Lhomond, 75231 Paris, France;\\
  \inst{2} Laboratoire de Physique des Solides, Univ. Paris-Sud, CNRS, UMR 8502, F-91405, Orsay, France;\\
  \inst{3} Physique et M\'ecanique des Milieux H\'et\'erog\`enes, IMR 7636, CNRS-UPMC-Paris 7-ESPCI, 75005 Paris, France;\\
  \inst{4} Laboratoire PCT, UMR Gulliver CNRS-ESPCI 7083, 10 Rue Vauquelin, 75231 Paris Cedex 05, France.
}

\abstract{
The behavior of weakly deformed drops on non wetting surfaces is
usually described using linear models. We show that these
simple pictures cannot account for measurements of the dynamics of
 droplets that oscillate or bounce on super-hydrophobic substrates.
We demonstrate that several peculiar experimental observations observed in previous works can be understood through a logarithmic correction of the linear model.
}

\begin{document}

%\title{Liquid Hertz Contact: Dynamics of weakly deformed drops on non wetting substrates}
%\author{F. Chevy}
%\author{F. Chevy}  \affiliation{Laboratoire Kastler-Brossel,
%ENS, 24 rue Lhomond, 75005 Paris}
%\author{A. Chepelianskii}
%\author{E. Rapha\"el}
%\date{\today}

%\pacs{03.75.Ss, 05.30.Fk, 32.80.Pj, 34.50.-s} \maketitle

\maketitle

Water impacts on surfaces are one the most common physical phenomena observed in every day life, with a wide range of applications, going from agriculture to inkjet printers.
%However, despite its high practical interest and a century long effort initiated by the seminal work of Worthington in the late nineteenth century \cite{worthington1876Forms}, a comprehensive description of impact phenomena is still missing. Nevertheless,
Recent progresses in experimental and theoretical methods have yielded a better understanding of this problem \cite{Rein1993Phenomena,yarin2006drop}. In particular, the development of novel superhydrophobic surfaces has allowed the study of these phenomena in well controlled situations \cite{richard2002surface,Lohse2012PRL}, where some of the most challenging effects, and most notably the line contact dynamics and hysteresis \cite{li2010dynamic}, can safely be neglected. %\cite{bergeron2001water}

In this Letter, we present a description of the dynamics of a sessile droplet in the regime of small deformation.  Contrasting with the commonly accepted scenario, we show that the droplet behaves as a non-linear spring whose stiffness decreases logarithmically with the deformation. Our model is based on an analogy with the classical Hertz contact for an elastic sphere and leads to accurate quantitative comparisons with existing data on superhydrophobic surfaces. We suggest that the analysis can be extended to the oscillations of droplets sitting on non-wetting substrates.

Consider a droplet of radius $R$, surface tension $\sigma$ and density $\rho$  at the surface of a non-wetting substrate. For $R\ll \kappa^{-1}$, where
$\kappa^{-1}=\sqrt{\sigma/\rho g}$ is the capillary length and $g$ is gravity, then the sag $\varepsilon$ the drop  (see Fig. \ref{Fig0}) is given by a balance between a linear capillary restoring
force $\simeq\sigma \varepsilon$ and the weight $\simeq\rho g R^3$ yielding the scaling relation $\varepsilon \sim \kappa^2 R^3$
\cite{mahadevan1999rolling,okumura2003water}.
%Including the coefficient in the argument, we expect, we expect in the limit of weak deformation $\kappa\varepsilon\simeq \kappa^2 R^2$.

\begin{figure}
\centerline{\includegraphics[width=\columnwidth]{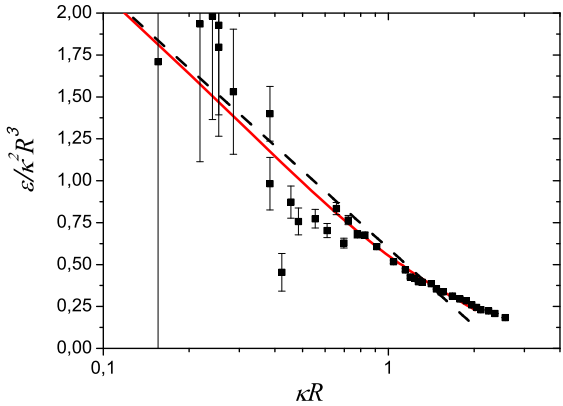}}
\caption{Color online. Measurement of the sag of the top of a liquid marble
obtained by deposition of lycopodium spores at the surface of a
water droplet. Black dashed line~: theoretical calculation using the numerical integration of the exact
Laplace law \ref{Eqn8}. Red solid line~: asymptotic expansion (Eq. \ref{Eqn2}) in the weakly
deformed regime. This asymptotic expansion is valid only in the regime $\kappa R\lesssim 1$ where the drop is weakly deformed.} \label{Fig1}
\end{figure}

\begin{figure}
\centerline{\includegraphics[width=\columnwidth]{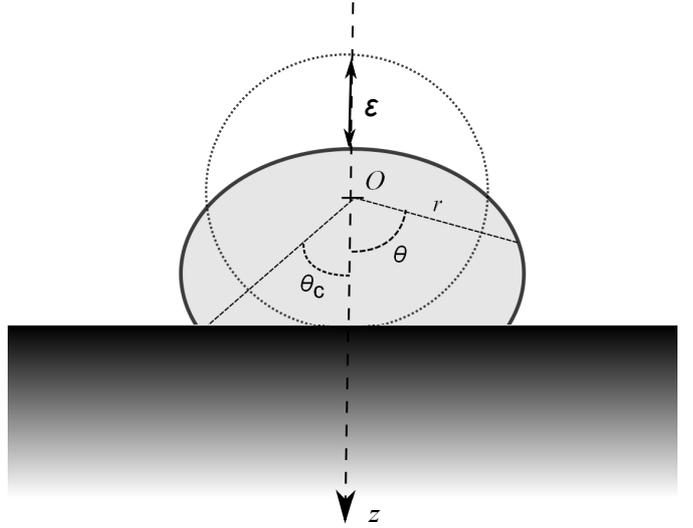}}
\caption{Solide circle: Schematic profile of the sessile droplet compared to the shape of a sphere of same  volume centered in O (dashed circle). $\theta_c$ represents the location of the ``contact" line of the drop with the surface. The sag of the drop is measured by the shift $\varepsilon$ of the top of the droplet compared to that of the unsquashed sphere.}
\label{Fig0}
\end{figure}

We measured the dimensionless sag $\varepsilon/\kappa^2 R^3$ for water droplets coated with lycopodium grains (a way to generate
superhydrophobicity \cite{aussillous2001liquid}), and the results are
presented in figure \ref{Fig1}. As seen on this semi-log plot, this ratio is not observed to be independent on the drop radius for $R\kappa<1$, but it weakly (logarithmically) diverges, suggesting rather a scaling

\be \varepsilon\sim \kappa^2R^3\left|\ln (\kappa R)\right|.
\label{Eqn1} \ee

This singular scaling was first pointed out in a study of the static properties of wet foams \cite{morse1993droplet}, and it can be justified by calculating the shape of the drop in the regime of weak deformation using an expansion very similar to that of \cite{shanahan82approximate,rienstra1990shape}.
Laplace equation relating the pressure discontinuity to
the mean curvature $C$ of the liquid/air interface reads in spherical coordinates (Fig. \ref{Fig0})

\be P'_0+\rho g r\cos\theta=P_0+2\sigma C,
\label{Eqn8}
\ee

\noindent where $P_0$ is the atmospherical pressure and $P'_0$ is an integration constant for the hydrostatic equations. In the limit of small deformations, we write $r=R+\zeta$, with
$\zeta\ll R$, and we obtain \cite{Landau59Fluid} for an axisymmetric drop

\be
\frac{\sigma}{R^2}\left(2+\Delta_{\theta}\right)\zeta=\Delta
P_0+\rho g R\cos\theta,
\label{Eqn10}
\ee

\noindent where $\Delta P_0=P'_0-P_0-2\sigma/R$ and $\Delta_{\theta}$ is the angular Laplacian given by

\be
\Delta_{\theta}=\frac{1}{\sin(\theta)}\partial_\theta\left(\sin\theta\partial_\theta(\cdot)\right). \ee

 This equation is supplemented by the following conditions: 1) Conservation of volume, $\int \sin\theta {\rm d}\theta\zeta=0$; 2) The position of the ``contact" at $\theta=\theta_c$ is given by $(R+\zeta (\theta_c))\cos\theta_c=R$ (see Fig. \ref{Fig0}); 3) The contact angle is 180$^\circ$, implying that $\zeta'(\theta_c)=R\tan\theta_c$.

These equations are readily solved and
 we get

\be \zeta(\theta)=\frac{\kappa^2
R^3}{3}\left[\frac{1+\cos \theta}{2}+\cos\theta\ln\left(\frac{1-\cos\theta}{1-\cos\theta_c}\right)\right],
 \ee

\noindent  with

\be
\theta^2_c\sim 2\kappa^2R^2/3,
\label{Eq:ContactAngle}
\ee
as derived in \cite{mahadevan1999rolling} and checked experimentally
in \cite{aussillous2001liquid}). We then calculate the static sag
of the top of the droplet
$\varepsilon=-\zeta(\pi)$, and that of the center of mass $\varepsilon_G$, yielding respectively:

\begin{eqnarray}
\varepsilon&=&\frac{\kappa^2R^3}{3}\left|\ln\left(\frac{\kappa^2R^2}{6}\right)\right|.
 \label{Eqn2}\\
\varepsilon_G&=&\frac{\kappa^2R^3}{3}\left|\ln\left(e^{5/6}\frac{\kappa^2R^2}{6}\right)\right|.
 \label{Eqn4}
  \end{eqnarray}

\noindent where we recover the logarithmic correction assumed in Eq. (\ref{Eqn1}). In order to interpret these equations, we invert Eq. \ref{Eqn4}, which yields to leading order:

\be
\frac{4\pi\rho R^3}{3}g=k_{\rm eff}\varepsilon_G,
\ee

\noindent with

\be
k_{\rm eff}=\frac{4\pi\sigma}{\left|\ln\left(e^{5/6}\varepsilon_G/2R\right)\right|}.
\label{Eq:ForceBalance}
\ee

Eq. \ref{Eq:ForceBalance} can be interpreted as a balance between the liquid weight of the droplet and a non-linear elastic force $k_{\rm eff}\varepsilon_G$, with an effective stiffness $k_{\rm eff}$. Remarkably, the stiffness vanishes for small deformations, meaning that the liquid spring becomes softer when the deformation is weaker. During small amplitude impacts or oscillations, we therefore expect the characteristic time of deformation  to diverge as $\sqrt{k_{\rm eff}}\sim\sqrt{\left|\ln\left(\varepsilon/R\right)\right|}$.

Building on this qualitative argument, we now give a more quantitative calculation of the dynamics of the droplet. Using the same argument as for Hertz contact \cite{hertz1881uber}, the existence of a characteristic timescale longer than the oscillation period $\sqrt{\rho g R^3/\sigma}$ of the Rayleigh oscillations of a free droplet allows one to use the static response studied in the previous section to calculate its dynamical properties. In the weak deformation limit the pressure in the liquid is given in first approximation by $2\sigma/R$, and Newton's law can be written as

\be
M\ddot\varepsilon_G=Mg-\frac{2\sigma}{R}\Sigma (t), \label{Eqn5}
\ee

\noindent where $M$ is the mass of the drop, and $\Sigma(t)=\pi R^2\theta_c^2 (t)$ is the ``contact" area. The problem is closed by deducing from Eq. (\ref{Eq:ContactAngle}) and (\ref{Eqn4}) the relationship between $\varepsilon_G$ and $\Sigma$

\be
\varepsilon_G=-\frac{\Sigma}{2\pi R}\ln\left(\frac{e^{5/6}\Sigma}{4\pi R^2}\right). \label{Eqn6}
\ee

\begin{figure}
%\begin{tabular}{c}
\centerline{\includegraphics[width=\columnwidth]{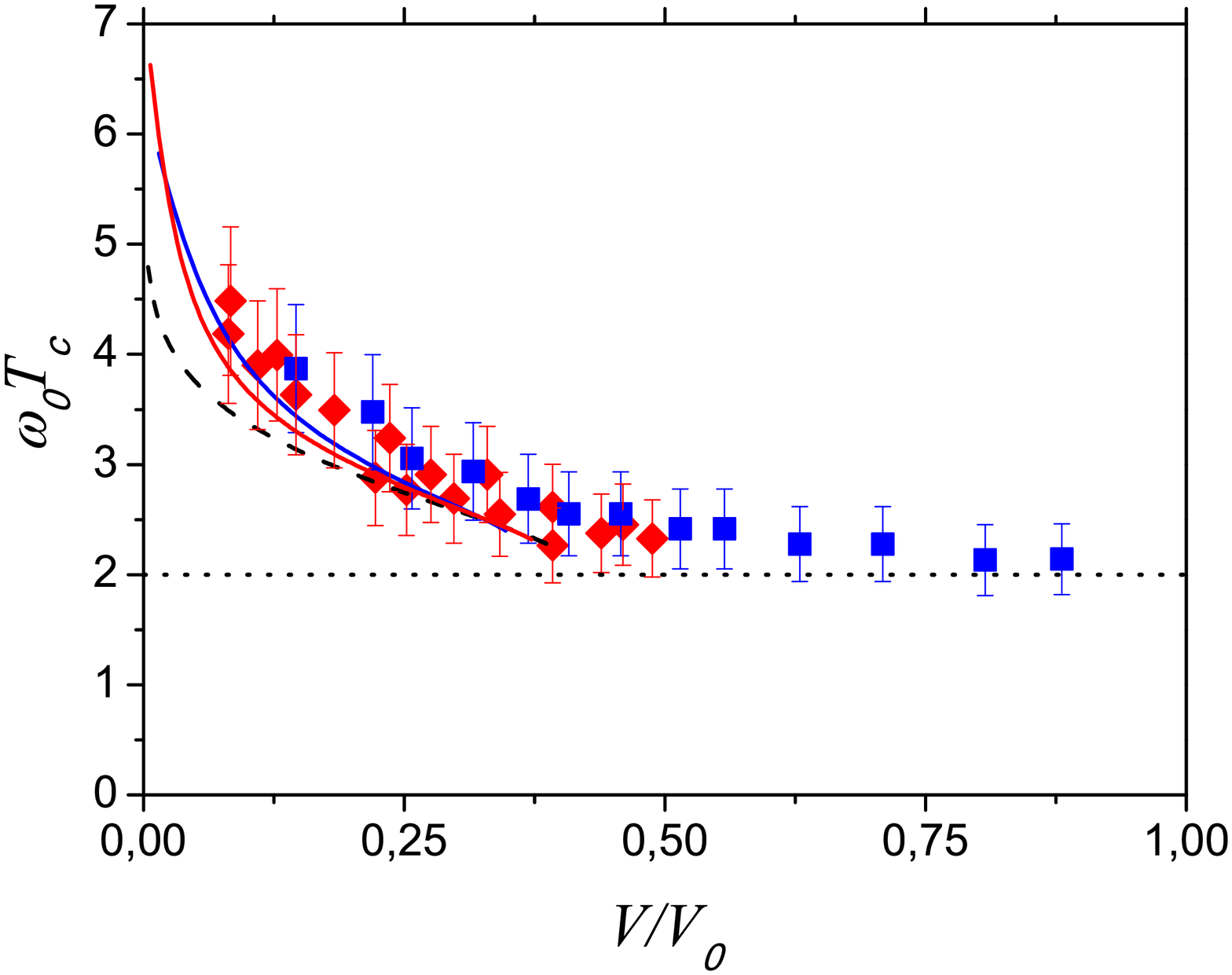}}
%\vskip -1.5cm
%\scalebox{0.7}{\includegraphics{Inverted.eps}}\\
%\vskip -1.5cm
\centerline{\includegraphics[width=\columnwidth]{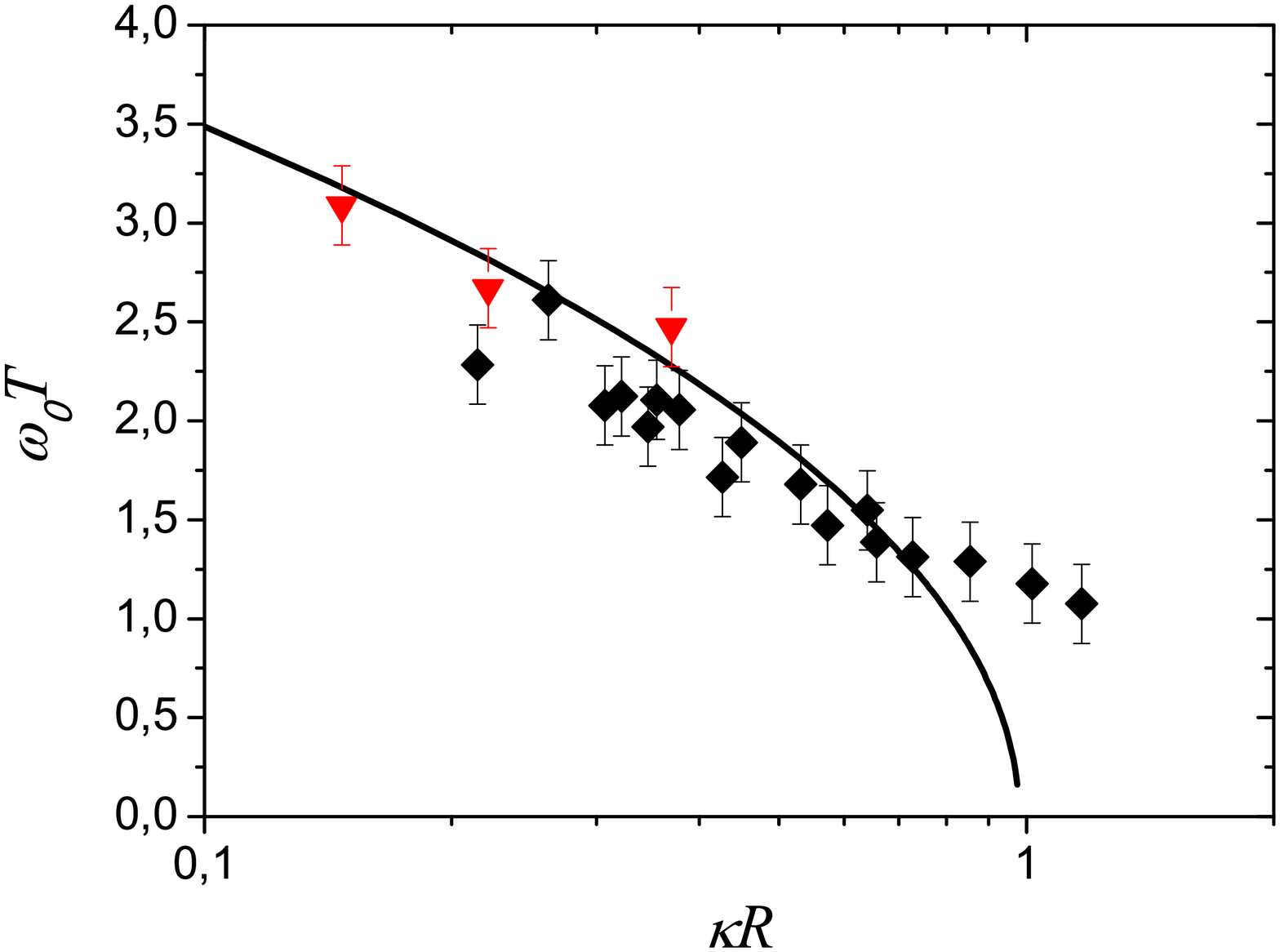}}
%\end{tabular}
\caption{Dynamics of a weakly deformed drop on a non wetting substrate. For all graphs, times are scaled in units of $1/\omega_0=\sqrt{\rho R^3/\sigma}$, velocities in units of $\sqrt{\sigma/\rho R}$, and lengths in units of $\kappa^{-1}=\sqrt{\sigma/\rho g}$. Top: Contact time during an impact.  Black squares: drops of radius 0.4~mm; Red diamonds: drops of radius 0.6~mm. Dashed line : prediction of (\ref{Eqn3}) without gravity; Black and red solid lines: prediction of Eq. (\ref{Eqn3}) with gravity for drops of radii 0.4~mm and 0.6~mm. Bottom: Small amplitude oscillation period of a sessile drop: black diamonds: resonance frequency of a mercury droplet; red triangles: free oscillation frequency of a coated water droplet; black line: asymptotic expansion (\ref{Eqn7}) for small radii. Note that for both graphs, theory is plotted without any adjustable parameter and that its range of validity is restricted to the limit of small deformation $V/V_0\ll 1$ and $\kappa R\ll 1$ respectively.}
\label{Fig2}
\end{figure}

Using Eq. (\ref{Eqn5}) and (\ref{Eqn6}), we first study the impact of the droplet bouncing off a non-wetting surface. Just before impact we have  $\varepsilon=0$, and $\dot\varepsilon=V$, where $V$ is the impact velocity. The contact duration $T_{\rm c}$ is calculated readily by noting that the right-hand term in Eq. (\ref{Eqn6}) derives from a potential energy $U(\varepsilon)$ defined by the implicit relation:

    $$dU=M\left(g-2\sigma\Sigma (\varepsilon_G)/R_0\right)d\varepsilon_G.$$

      Integrating the associated conservation of energy then leads to

    \be
    T_{\rm c}=\int_0^{\varepsilon_m}\frac{2d\varepsilon}{\sqrt{V^2-2U(\varepsilon)/M}},
    \label{Eqn3}
    \ee

\noindent where $\varepsilon_m$ is the maximum deformation of the drop, defined by $U(\varepsilon_m)=MV^2/2$. In Fig. \ref{Fig2} we compare the prediction of equation (\ref{Eqn3}) with data of ref. \cite{okumura2003water}) obtained for droplets of radii $R=0.4$~mm and $R=0.6$~mm bouncing over a superhydrophobic surface. The fairly good agreement between experimental points  and theory provides a first validation of our approximations. In particular the model explains why the contact time increases in the regime of small deformations ($V$ small) where we may naively have expected a constant contact time.
%Our model predicts that deviations should occur for $\varepsilon\ll R$, that is, for ${\rm We}=\rho V^2 R/\sigma$ smaller than unity, as observed in Fig. \ref{Fig2}.a).

Moreover, Eq. (\ref{Eqn5}) and (\ref{Eqn6}) allow one to study the small amplitude oscillations of the drop around equilibrium. In this pursuit, we linearize Eq. (\ref{Eqn5}), which can be written as $\ddot\varepsilon_G=-\omega^2\varepsilon_G$, with

\begin{equation}
\omega^2=-\frac{3\sigma}{2\pi \rho R^4}\left(\frac{d\Sigma}{d\varepsilon_G}\right)_0,
\label{Eqn9}
\end{equation}

\noindent where the subscript 0 indicates that the derivative is evaluated at equilibrium. Differentiating Eq. (\ref{Eqn6}), one obtains after some straightforward algebra,

\begin{equation}
\omega^2=\frac{3\sigma}{\rho R^3}\left|\frac{1}{\ln\left(\frac{e^{11/6}}{6}\kappa^2R^2\right)}\right| \label{Eqn7}
\end{equation}

To test experimentally this prediction, we measured the free oscillation period of Hg droplets as well as water droplets on superhydrophobic surfaces. The results are displayed in \ref{Fig2}.b) as a function of the drop radius. It is observed that the dimensionless period is not constant, but increases (by a factor 3) as the droplet radius decreases. Again the system behaves in a softer way when it is more weakly deformed. In this limit, Eq. \ref{Eqn7}  compares favorably with the data, without any adjustable parameters, except in the limit $\kappa R\gtrsim 1$ for which the drop is strongly flattened by gravity and cannot be described within a weak deformation approximation.

The later result contradicts an earlier work on the oscillations of sessile drops where no logarithmic behavior was found  \cite{courty2006oscillating}. We argue that this discrepancy is due to an inconsistency in the calculation presented in \cite{courty2006oscillating} where the eigenmodes of the droplet were decomposed on a single spherical harmonic even though  the system is no longer rotationally invariant in the presence of the substrate. We show here that reintroducing the full set of spherical harmonics allows us to recover Eq. (\ref{Eqn7}) and extends it to higher order modes.

For high frequency modes, the quasi-static approximation performed above is no longer valid, and one needs in principle to solve the full Rayleigh-like problem in the presence of boundary conditions imposed by the substrate. In \cite{courty2006oscillating}, it was proposed to replace the boundary condition by a pressure field $P_{\rm c}(\bm\Omega)$ located on the contact region. In the limit of small droplets, the contact area is small and the pressure field can be approximated by a Dirac distribution

\be
P_{\rm c}(\bm\Omega)=\frac{F_c}{R^2} \delta(\bm\Omega)=\frac{F_c}{R^2}\sum_{\ell=0}^\infty Y_{\ell}^0(\theta)Y_{\ell}^0(\theta=0)^*,
\ee

\noindent where $F_c$ is the force exerted by the substrate on the droplet and where we have decomposed the angular Dirac $\delta$ on the basis of spherical harmonic $Y_\ell^m$ (by symmetry around the $z$ axis, the $m\not =0$ do not contribute).  Using the definition of the spherical harmonics, we have $Y_{\ell}^0(\theta=0)=\sqrt{(2\ell+1)/4\pi}$, hence

\be
P_{\rm c}(\bm\Omega)=\frac{F_c}{3R^2}\sum_{\ell=0}^\infty \sqrt{\frac{2\ell+1}{4\pi}}Y_{\ell}^0(\theta),
\label{Eqn12}
\ee

The presence of this pressure field modifies the Laplace condition, and we now have

\be
P_{\rm c}+\frac{\sigma}{R^2}\left(2+\Delta_{\theta,\varphi}\right)\zeta=\Delta
P_0+\rho g R\cos\theta,
\label{Eqn10}
\ee

This equation mixes the effects of gravity and oscillations. However we can decouple them using the linearity of Eq. \ref{Eqn10}. We thus set $\zeta=\zeta^{(0)}+\zeta^{(1)}$ and $F_c=F_c^{(0)}+F_c^{(1)}$, where the subscript 0 (1) denotes the static (dynamic) part of the deformation. Let's now project this equation on the basis of the spherical harmonics. Since $P_c$ possesses non-zero components on all $Y_\ell^0$,  $\zeta^{(1)}$ must be written as $\zeta^{(1)}(\bm\Omega)=\sum_\ell \zeta^{(1)}_\ell(t) Y_\ell^0(\bm\Omega)$. This point is the main difference with the analysis of \cite{courty2006oscillating} where a less general form $\zeta^{(1)}=\zeta_\ell^{(1)}(t)Y_\ell^0 (\Omega)$ was assumed, in contradiction with the decomposition of $P_c$ on the full set of $\ell$ values.

In the case of an inviscid fluid, the flow in the bulk can be derived from a velocity potential $\psi(\bm r,t)$ solution of the Laplace equation $\Delta\psi=0$. Using the boundary conditions at the surface of the droplet $\partial_r\psi(r=R,\bm\Omega)=\partial_t\zeta (\bm\Omega)$, the decomposition of $\psi$ of the spherical harmonic can be readily expressed in terms of the $\zeta{(1)}_\ell$'s and we have

\be
\psi(r,\bm\Omega,t)=\sum_{\ell}\dot\zeta^{(1)}_\ell\left(\frac{r^\ell}{\ell R^{\ell-1}}\right)Y_\ell^0(\bm\Omega).
\ee

Using the time-dependent Bernoulli equation $\rho\partial_t\psi=-p$, we can express the pressure field  in Eq. (\ref{Eqn10}) in terms of $\zeta_\ell$. Projecting Eq. (\ref{Eqn10}) on the basis of the spherical harmonics and using the condition $\Delta_{\theta,\varphi}Y_\ell^m=-\ell(\ell+1)Y_\ell^m$, we get an expression for $\zeta_\ell$:

\be
\zeta_\ell^{(1)}=\frac{F_c}{\sigma}\frac{\ell}{\ell(\ell-1)(\ell+2)-\omega^2/\omega_0^2}\sqrt{\frac{2\ell+1}{4\pi}}
\label{Eqn11}
\ee

\noindent with $\omega_0^2=\sigma/\rho R^3$. To obtain the excitation spectrum of the droplet, we use the condition $\zeta^{(1)} (\theta=0)=0$ imposed by the substrate. With the expression of $\zeta^{(1)}_\ell$ obtained above, this yields the following equation for the eigenfrequency $\omega$

\be
\sum_{\ell=1}^\infty\frac{\ell(2\ell+1)}{\omega^2/\omega_0^2-\ell(\ell-1)(\ell+2)}=0,
\label{Eqn13}
\ee

The sum  in Eq. (\ref{Eqn13}) is divergent and it must be regularized. The origin of this divergence comes from the singular nature of the $\delta$ pressure-field used to describe the contact between the substrate and the droplet. Indeed, modes with high values of $\ell$ oscillate very fast spatially (on angular ranges of the order $\theta\simeq 2\pi/\ell$) and can therefore be sensitive to the actual shape of the contact region. This suggests to introduce in the sum (\ref{Eqn13}) a cutoff at $\ell_{\rm max}\simeq \theta_c^{-1}\simeq(\kappa R)^{-1}$, as done in  the static case. Using this assumption, we reformulate Eq. (\ref{Eqn13}) as

\be
\sum_{\ell=0}^{\ell_{\rm max}}\frac{\ell(2\ell+1)}{\omega^2/\omega_0^2-\ell(\ell-1)(\ell+2)}=0.
\label{Eqn13b}
\ee

To get a further insight on the solutions of this equation, we recast it as

\be
\sum_{\ell=1}^{\ell_{\rm max}}\left[\frac{\ell(2\ell+1)}{\omega^2/\omega_0^2-\ell(\ell-1)(\ell+2)}+\frac{2}{\ell}\right]=2\sum_{\ell=1}^{\ell_{\rm max}}\frac{1}{\ell},
\label{Eqn14}
\ee

The series in the lhs of the equation is now convergent, and we can safely take the limit $\ell_{\rm max}\rightarrow\infty$. As for the rhs, using the definition of the Euler constant $\gamma$, it can be expanded as $\sum_{\ell=1}^{\ell_{\rm max}}\ell^{-1}=\ln(\ell_{\rm max})+\gamma+o(1)$. Finally, we can formally rewrite Eq. (\ref{Eqn14}) as

\be
F(\omega^2/\omega_0^2)=2\left(\ln(\ell_{\rm max})+\gamma\right),
\label{Eqn15}
\ee

\noindent with $F$ being defined by

\be
F(z)=\sum_{\ell=1}^\infty\left[\frac{\ell(2\ell+1)}{z-\ell(\ell-1)(\ell+2)}+\frac{2}{\ell}\right].
\ee

Since $\ell_{\rm max}$ is large, the solutions of Eq. (\ref{Eqn15}) must be close to the poles of $F$ located at the values $z_\ell=\ell (\ell-1)(\ell+2)$ corresponding to the Rayleigh frequencies of a levitated droplet. Let us expand $F(z)$ close to $z_\ell$ up to the constant terms. We have in this limit $F(z)\simeq \ell(2\ell+1)/(z-z_\ell)+a_\ell+o(1)$, where $a_\ell$ is a number that can be calculated numerically ($a_1=-1/6$, $a_2=2.0889$, $a_3=3.0927$,...). To this approximation we get the following dispersion relation for the sessile droplet

\be
\frac{\omega_\ell^2}{\omega_0^2}=\ell(\ell-1)(\ell+2)+\frac{\ell (2\ell+1)}{\ln\left(\ell_{\rm max}^2 e^{2\gamma-a_\ell}\right)}.
\label{Eqn16}
\ee

Except for the case $\ell=1$ for which $z_1=0$, the eigenfrequencies of the sessile droplet are given by the Rayleigh frequencies of a levitated droplet, up to a logarithmic correction. The case $\ell=1$ corresponds to the low frequency mode studied in the first part of this Letter. Eq. (\ref{Eqn7}) and (\ref{Eqn16}) coincide if $\ell_{\rm max}=\sqrt{6}e^{-1-\gamma}/\kappa R$, which agrees with the general scaling introduced earlier to define $\ell_{\rm max}$.

\begin{figure}
\centerline{\includegraphics[width=\columnwidth]{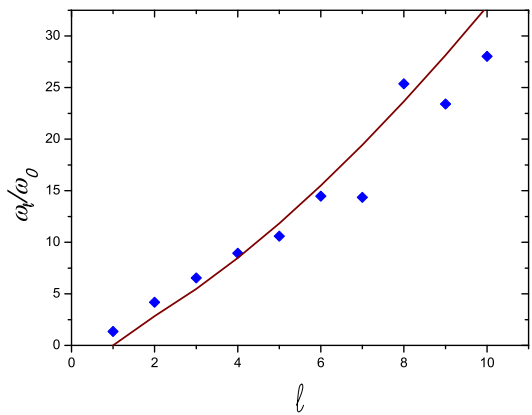}}
\caption{Comparison between resonances of vibrated Mercury droplets from \cite{courty2006oscillating}, and the Rayleigh prediction $\omega_\ell/\omega_0=\sqrt{\ell(\ell-1)(\ell+2)}$.}
\label{Fig3}
\end{figure}

We now compare this prediction to the measurements of ref. \cite{courty2006oscillating}. We argue that in this article the values of $\ell$ were not  attributed to the proper modes. Indeed, Courty {\em et al.} claimed that, by monitoring the center of mass of the droplet excited by the vibration of the substrate along the vertical direction, their detection scheme could only detect modes with even values of $\ell$ and thus attributed the value $\ell=2n$ to the $n$-th mode. On the contrary, we argue here that the apparatus could track all resonances, since all modes are coupled to the $\ell=1$ spherical harmonic describing the motion of the center of mass. We thus attribute here the value $\ell=n$ to the $n$-th resonance observed experimentally. Using this new attribution, we observe a fairly good agreement between the position of the resonances observed by Courty {\em et al.} and the prediction of Eq. (\ref{Eqn16}) (see Fig. \ref{Fig3})

The comparison between the predictions of Eq. (\ref{Eqn5}) and (\ref{Eqn6}) with existing experimental data support our model of logarithmic spring to describe the dynamics of a weakly deformed droplet. In future works, we plan to extend our findings to other experimental situations which are yet unresolved. Among these, we will include hysteresis of the contact angle observed on real surfaces, which might explain the critical velocity below which the drop does not bounce anymore \cite{Crick2011}. We will also study the impact on elastic or liquid surfaces. This would for instance help understanding recent experiments on droplets walking at the surface of water \cite{couder2005walking}. Finally, it could be interesting to extend the present analysis to higher deformations to probe impacts at high velocity and explain the saturation of the contact time observed in \cite{richard2002surface}.

FC acknowledges support from Institut Universitaire de France. We thank C. Clanet and G. Lagubeau for helpful discussions and P. Aussillous for providing us with the data of Fig. \ref{Fig1}.

\bibliographystyle{unsrt}
\bibliography{BibHydro}

\end{document}